\documentclass[12pt]{iopart}
\usepackage{graphicx}
\usepackage{overpic}
\usepackage{amssymb}
\newcommand{\degree}{$^{\circ}$}
\begin{document}
\title[Bulk graphanes synthesized from benzene and pyridine]{Bulk graphanes synthesized from benzene and pyridine}
\author{M.V. Kondrin $^1$, N.A. Nikolaev $^1$,  K.N. Boldyrev $^2$ , Y.M. Shulga $^3$ , I.P. Zibrov $^1$, V.V. Brazhkin $^1$ }
\address{ $^1$ Institute for High Pressure Physics RAS, 142190 Troitsk, Moscow, Russia}
\address{ $^2$ Institute of Spectroscopy RAS, 142190 Troitsk, Moscow, Russia}
\address{ $^3$ National University of Science and Technology 'MISIS', Leninsky Pr. 4, 119049 Moscow, Russia}
\ead{mkondrin@hppi.troitsk.ru}

\begin{abstract}
The discovery of graphene stimulated an intensive search for its analogs and derivatives. One of the most interesting derivatives is hydrographene called graphane. Calculations indicate that bulk graphane is thermodynamically more favorable than all CH hydrocarbons including benzene. At the same time, pressure-induced polymerization of hydrocarbons and their derivatives at room temperature leads to the formation of amorphous products or poorly ordered one-dimensional products such as polyacetylene and benzene-derived carbon nanothreads. 

Here, we report a high-pressure high-temperature synthesis of several millimeter-sized samples of bulk graphanes with the composition C--H(D) from benzene and graphene-derivative C--H--N$_{0.2}$ from pyridine. X-ray diffraction, transmission electron microscopy, and infrared spectroscopy of new materials reveal relatively large (several nanometers in size) crystalline grains of an sp$^3$-bonded graphane lattice (3-cycle-4-step, the orthorhombic structure with $Pbca$ space-group and parameters $a$~=~9.5--9.8~\AA, $b$~=~8.9--9.1~\AA, $c$~=~17.1--17.3~\AA). The main hydrogen groups in samples are C-H groups connected by aliphatic bonds. The synthesized graphanes at atmospheric pressure are stable up to 500\degree~C. The macroscopic density of CH samples is 1.5--1.57~g~cm$^{-3}$ and the refractive index is 1.78--1.80. The absorption spectra of samples with a high degree of crystallization exhibits a weak absorption maximum at 2.8~eV, which is responsible for the yellow-orange color, large absorption maximum at 4~eV and an absorption edge associated with the width of the optical gap at 5.2~eV. The bulk modulus (30--37~GPa) and shear modulus (15--18~GPa) of the fabricated samples, as well as their hardness (1--1.5~GPa), are about twice as high as the respective values for polycrystalline graphite. The solution of metalorganic complexes in benzene and pyridine makes it possible to obtain doped graphanes, which can have extraordinary electron transport and magnetic properties.
\end{abstract}
\maketitle
\section{Introduction}
The discovery of graphene\cite{geim:nm07} stimulated an intensive theoretical and experimental search for chemically modified carbon sheets\cite{dreyer:acie10}. Attention is primarily focused on the possibility of the synthesis of graphane, i.e., a fully saturated CH sheet, four-coordinated at C \cite{sofo:prb07,sluiter:prb03,elias:s09}. This material was obtained at the nanometer scale, but both the synthesis and identification are still problematic. In addition to the interest in quasi-two-dimensional graphane, researchers focused attention on the potential possibility of the synthesis of three-dimensional multilayer graphane crystals\cite{wen:pnas11,artyukhov:jpca10,rohrer:prb11,wen:jacs11,sahin:wir15,zhou:nrl14}. It was predicted that bulk graphane is thermodynamically much more stable than benzene, which is considered in classical chemistry as the most stable C-H hydrocarbon\cite{wen:pnas11,sahin:wir15,zhou:nrl14,wen:jacs11}. Bulk graphanes and their doped derivatives can be used for hydrogen storage, as new promising magnetic, superconducting, and optical materials\cite{savini:prl10,sahin:wir15,zhou:nrl14}. 	

One of the approaches to the synthesis of graphanes is the hydrogenation of graphenes and other nanostructured carbon materials\cite{sofer:n12,sofer:n14,bashkin:jetpl04}. In most cases, amorphous materials with the hydrogen content lower than the necessary value were obtained. An alternative method for obtaining graphanes involves initial hydrocarbons with initially correct C--H stoichiometry (acetylene, styrene, benzene), because calculations indicate that all these hydrocarbons are less stables than graphanes. The polymerization of such materials at moderate pressures results in the formation of quasi-one-dimensional polymer structures with a low degree of crystallization because of hindered kinetics and steric constraints \cite{kovacic:joc63}. At high pressures, the polymerization of acetylene is very intense even at room temperature \cite{ceppatelli:jcp00}. The properties of benzene and its derivative, pyridine C$_5$H$_5$N, at high pressures were studied in \cite{pruzan:jcp90,gauthier:hpr92,cansell:jcp93,ciabini:jcp02,schetino:csr07,citroni:pnas08,ciabini:nm07,bini14,zhuravlev:prb10,fanetti:jcp11,shinozaki:jcp14}. Most studies are in situ Raman and structural investigations. At pressures above 20-30 GPa at room temperature and at lower pressures at a higher temperature, benzene undergoes irreversible polymerization with the formation of amorphous modifications\cite{pruzan:jcp90,gauthier:hpr92,cansell:jcp93,ciabini:jcp02,schetino:csr07,citroni:pnas08,ciabini:nm07,bini14}. Pyridine, which is an analog of benzene, exhibits a similar behavior \cite{bini14, zhuravlev:prb10,fanetti:jcp11} as well as other derivatives of benzene like furan\cite{ceppatelli:jcp03} and thiophene\cite{pruzan:jcp92}. It has recently been shown that an increase in the high-pressure treatment time provides partially ordered quasi-one-dimensional nanothreads immersed into amorphous matrix\cite{fitzgibbons:nm14}. 
 
In this work, to simplify the kinetics of crystallization of polymers formed from benzene and pyridine, we used the processing of molecular materials at high pressures (6--10~GPa) and high temperatures (700--1200~K). Attention is primarily focused on the synthesis and investigation of samples obtained by the thermobaric treatment of benzene.

\section{Methods.}

We used benzene (purity 99.5\%), deuterated benzene (purity 95\%), pyridine (purity 99.0\%), ferrocene (purity 99.5\%), and butyllithium (purity 95\%) as the initial substances.

The experiments were performed at calibrated CONAC-28 apparatuses at high pressures up to 8 GPa and CONAC-15 apparatuses at pressures up to 10 GPa. The initial liquids were filled into hermetic copper or platinum capsules in an argon box. Heating was ensured by a current flowing through an external graphite heater isolated from capsules. The heating and cooling rate was 10 K min$^{-1}$. The aging time at the maximum temperature was 1 h.  Samples 5 mm in diameter and 3.5 mm in height (50--70 mg) were synthesized at pressures of 6--8 GPa and temperatures of 700--1200 K; samples 2 mm both in diameter and in height (10--15 mg) were obtained at pressures of 8--10 GPa). 

The elemental C, H (by the combustion method) analysis of the synthesized samples was conducted with the use of an Elementar CHNS/O Vario Micro cube analyzer.

The macroscopic density of the samples was measured by the sinking method. The refractive index was measured by the immersion technique.

An AXS (Bruker) powder diffractometer with a two-dimensional detector ($^{Cu}$K$_{\alpha}$ radiation, Si monochromator, transmission mode) was used for the phase analysis of the samples.

For transmission electron microscopy studies, small amounts of powders were grinded in ethanol. The resulting suspension was deposited on a copper grid covered by a perforated carbon film. Electron diffraction and X-ray spectra were obtained from fragments of crystals in a JEOL JEM-2100 microscope (accelerating voltage 200 kV, resolution on the grating 0.14 nm) with an EDX analyzer.

Infrared spectra were recorded using a Perkin Elmer Spectrum 100 Fourier spectrometer equipped with UATR in the range from 675 to 4000 cm$^{-1}$. 

Transmission spectra were collected in the spectral range 190--1100 nm with the resolution 2 nm using Specord 250 Analytik Jena spectrometer. Measurements were carried out on the powder of the compound is placed in a quartz cuvette with an ethanol millimeter thickness. The absorption egde of ethanol is 6.5 eV. The transmission spectra of the sample divided by the cell with the alcohol spectrum (black curves). During the measurement of the sample particles gradually settled to the bottom of the cell, so the 3 sets was measured. Thus, numbers 1, 2 and 3 indicate the ordinal number of measurement (1 - minimum sediment, 3 - sediment maximum). The sample is not completely settled until the end of the last measurement.

The high-temperature stability of polymerized benzene and pyridine at normal pressure was studied in the atmosphere of pure argon (99.997\%) using a Netzsch STA 409 PC Luxx simultaneous thermal analyzer connected with a QMS 403C A\"{o}los quadrupole mass spectrometer. The temperature of the sample was increased up to 1170 K with a heating rate of 10 K min$^{-1}$. All emitted gases and vapors were transported from the sample area to the ion source of the mass spectrometer for analysis. 
 
The hardness was measured by the Vickers method with the 1N load. The elastic properties were studied by measuring the longitudinal and transverse velocities of 10-MHz ultrasound passing through a plane-parallel sample at an original setup \cite{stalgorova:iet96}. 

\section{Results and discussion.}

\begin{figure}
\includegraphics[width=0.7\textwidth]{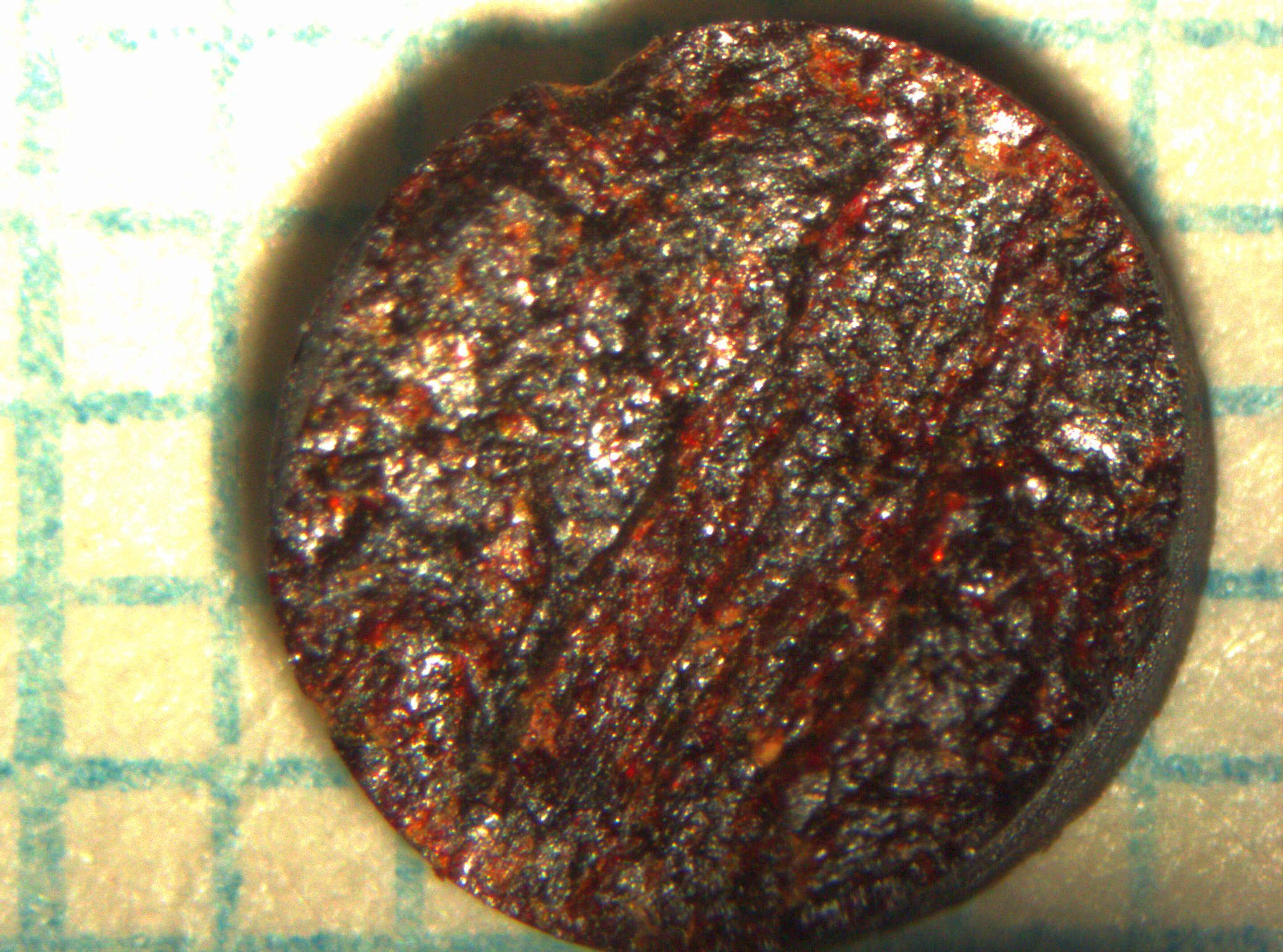}\\
\includegraphics[width=0.7\textwidth]{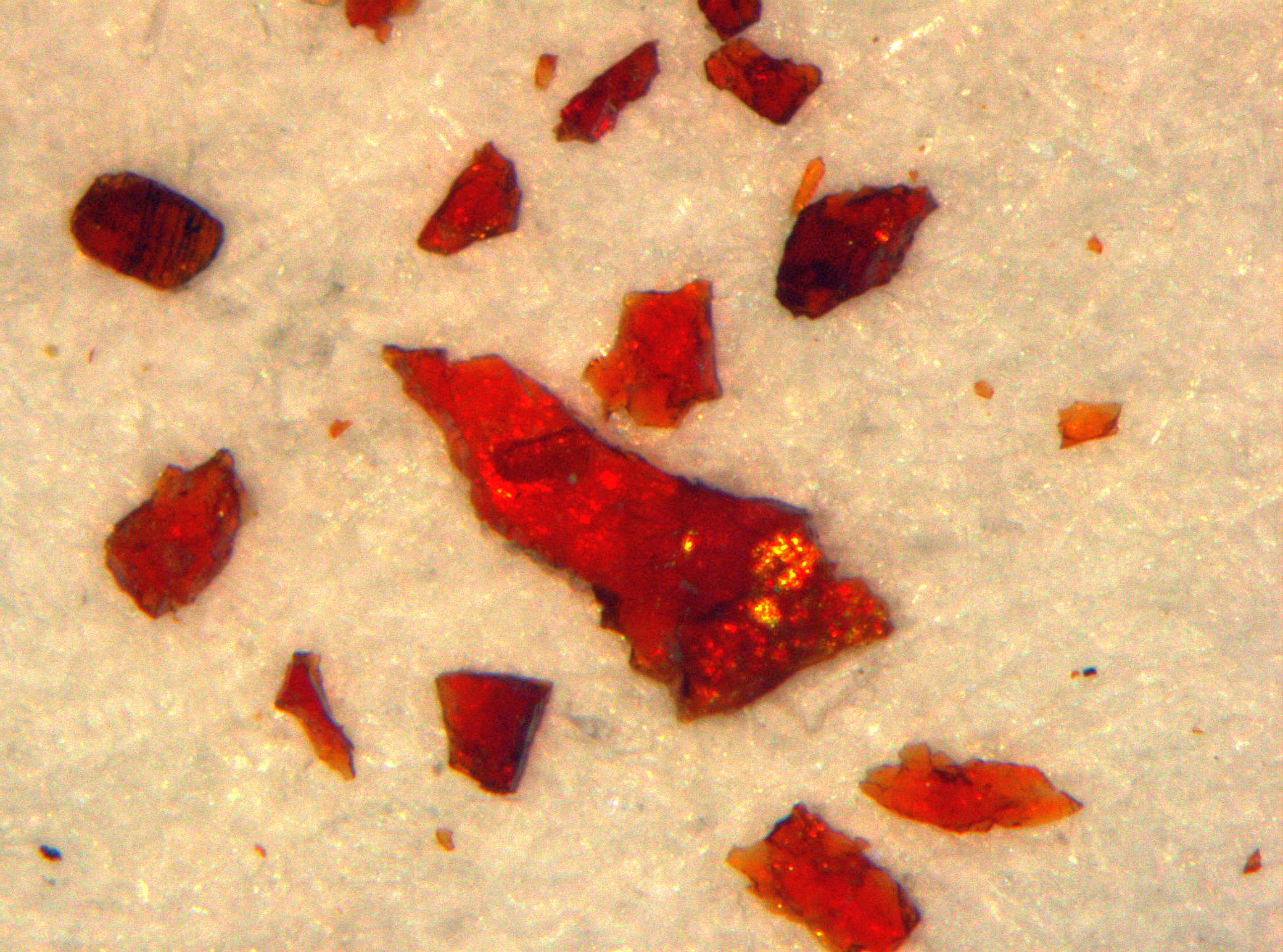}
\caption{Typical views of freshly synthesized sample (with $\oslash$ 5.5 mm and 5.5 mg wheight) and small splinters ( $\approx 0.2$ mm in length) of it when broken. }
\label{photo}
\end{figure}

The high-pressure high-temperature treatment of benzene, pyridine, and their solutions was performed in hermetic copper or platinum capsules with the use of CONAC apparatuses. Samples synthesized at pressures above 7~GPa and temperatures of 900--1000~K had the yellow-orange color (Fig.~\ref{photo}), but samples obtained at lower pressures were almost colorless. All samples were dielectric and their thin layers were transparent.

\begin{figure}
\includegraphics[width=0.7\textwidth]{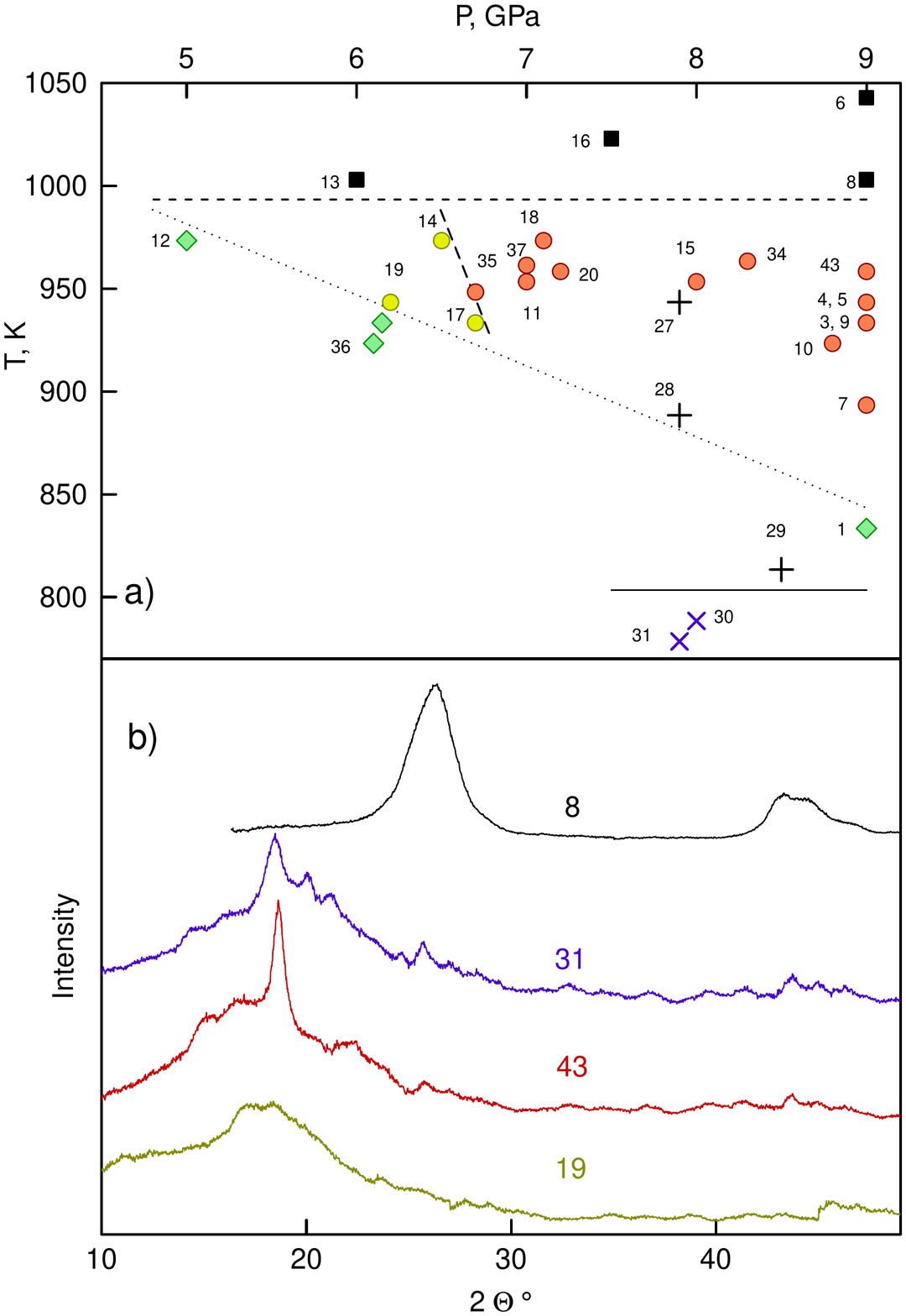}
\caption{a) P-T diagram of benzene (both C$_6$H$_6$ and C$_6$D$_6$) and pyridine transformations in vicinity of polymerization region. $\square$ and $+$ -- nanographite phases obtained from benzene and pyridine respectively, $\circ$ -- polymeric benzene phases (yellow -- ``soft'' low density and red -- ``hard'' high-density ones), $\times$ -- polymeric pyridine phases, $\lozenge$ -- benzene phases recovered in liquid form after pressure release. Lines stand for approximate transitions borders. b)--raw XRD patterns taken from different samples synthesized at different P-T conditions. Numbers beside curves and their colors correspond to points in panel a): 8 -- graphitic sample, 31 -- polymeric pyridine, 43 -- partially crystallized graphane, 19 -- amorphous hydrocarbon phase.}
\label{pt}
\end{figure}

We performed about 50 experiments on the synthesis of samples; the results were used to plot the phase diagram of transformations (Fig.~\ref{pt}~a). It is seen that the reduction of the synthesis pressure is accompanied by the narrowing of the temperature range in which polymer phases can be obtained, and benzene under heating at pressures below 6 GPa loses hydrogen without an intermediate polymerization phase. Samples with a high degree of crystallization can be obtained from benzene only at pressures above 8 GPa in the temperature range of 930--1000~K. The (P,T) regions of transformation of benzene to the amorphous and crystalline phases established in this work are in satisfactorily agreement with the data obtained in situ in diamond anvil cells and reported in \cite{cansell:jcp93,ciabini:jcp02,ciabini:nm07}. Bright orange samples with a high degree of crystallization that were obtained in our work at maximum pressures were possibly marked in \cite{cansell:jcp93} as ``polymer1'' and colorless amorphous phases that we obtained at pressures of 6--7.5~GPa, as ``polymer2''. Polymerized, partially crystallized pyridine-based phases at these pressures can be obtained in a temperature range of 750--800~K. A lower polymerization temperature of pyridine as compared to benzene correlates with lower polymerization pressures of pyridine at room temperature\cite{zhuravlev:prb10, fanetti:jcp11}.

The X-ray diffraction study showed that samples obtained from benzene and pyridine at temperatures below 1000 and 800~K, respectively, had an amorphous or amorphous-crystalline structure with different degrees of crystallization depending on the aging time and (P,T) conditions (Fig.~\ref{pt}~b). The elemental analysis shows that the initial stoichiometry in all these samples held with an accuracy of 2--3\%. Thus, it can be assumed that solid modifications obtained are polymerized phases based on benzene and pyridine. At higher treatment temperatures under pressure, samples lost hydrogen (hydrogen and nitrogen in the case of pyridine) and were transformed to an amorphous or nanocrystalline graphite-like material. 
	
\begin{figure}
\includegraphics[width=0.47\textwidth]{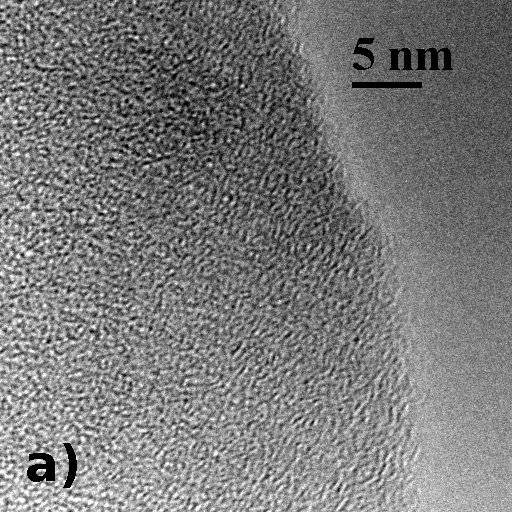}
\includegraphics[width=0.47\textwidth]{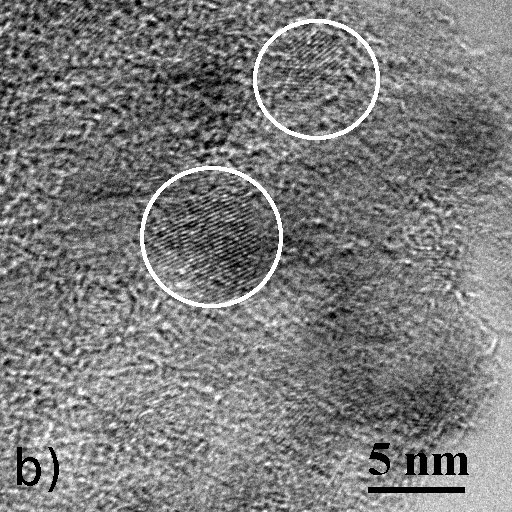}\\
\includegraphics[width=0.47\textwidth]{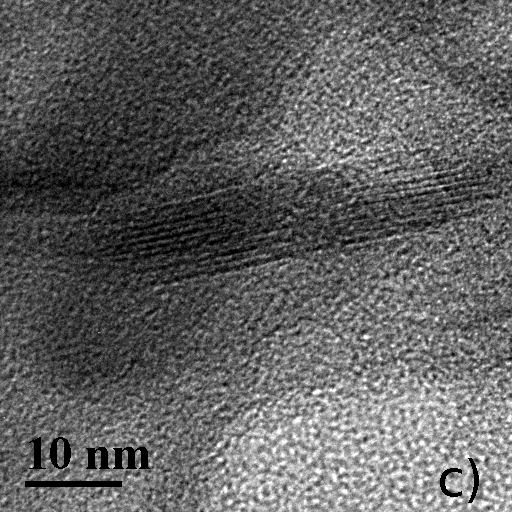}
\includegraphics[width=0.47\textwidth]{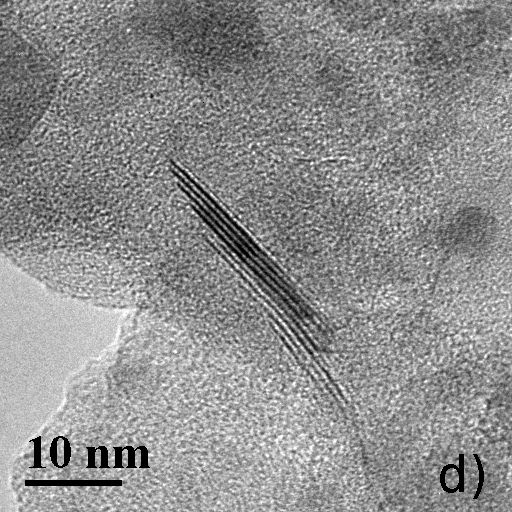}\\
\caption{Transmisson elecrton microscopy of samples showing different type of crystallized areas. a) -- crystallization onset,  b), c) -- different grains of graphane phase (marked by circles in panel b), d) -- nanothreads. Typical interatomic distances in pictures b)--d) are equal to 2.1, 3.6 and 6.5 \AA\ respectively. }
\label{tem}
\end{figure}

The transmission electron microscopy study of polymer modifications really indicates that samples is an amorphous matrix with a noticeable fraction of crystalline inclusions with a relatively large size of 2--4~nm (Fig.~\ref{tem}). In addition to well-crystallized regions (Fig.~\ref{tem}~b)-c)), there are both regions at the initial crystallization ordering stage (Fig.~\ref{tem}~a)) and rare quasi-one-dimensional inclusions similar to nanothreads obtained in \cite{fitzgibbons:nm14} (Fig.~\ref{tem}~d).

The density of amorphous modifications obtained in this work was 1.35--1.4~g~cm$^{-3}$, whereas the density of the partially crystallized samples was 1.5--1.57~g~cm$^{-3}$. The refractive index for partially crystallized samples was in the range of 1.78--1.80. It is noteworthy that amorphous benzene-based polymers obtained in the previous works had a density of 1.4~g~cm$^{-3 }$ and a refractive index of 1.75--1.76; both these quantities were determined with a large error \cite{pruzan:jcp90,gauthier:hpr92,bini14}. 

\begin{figure}
\includegraphics[width=0.8\textwidth]{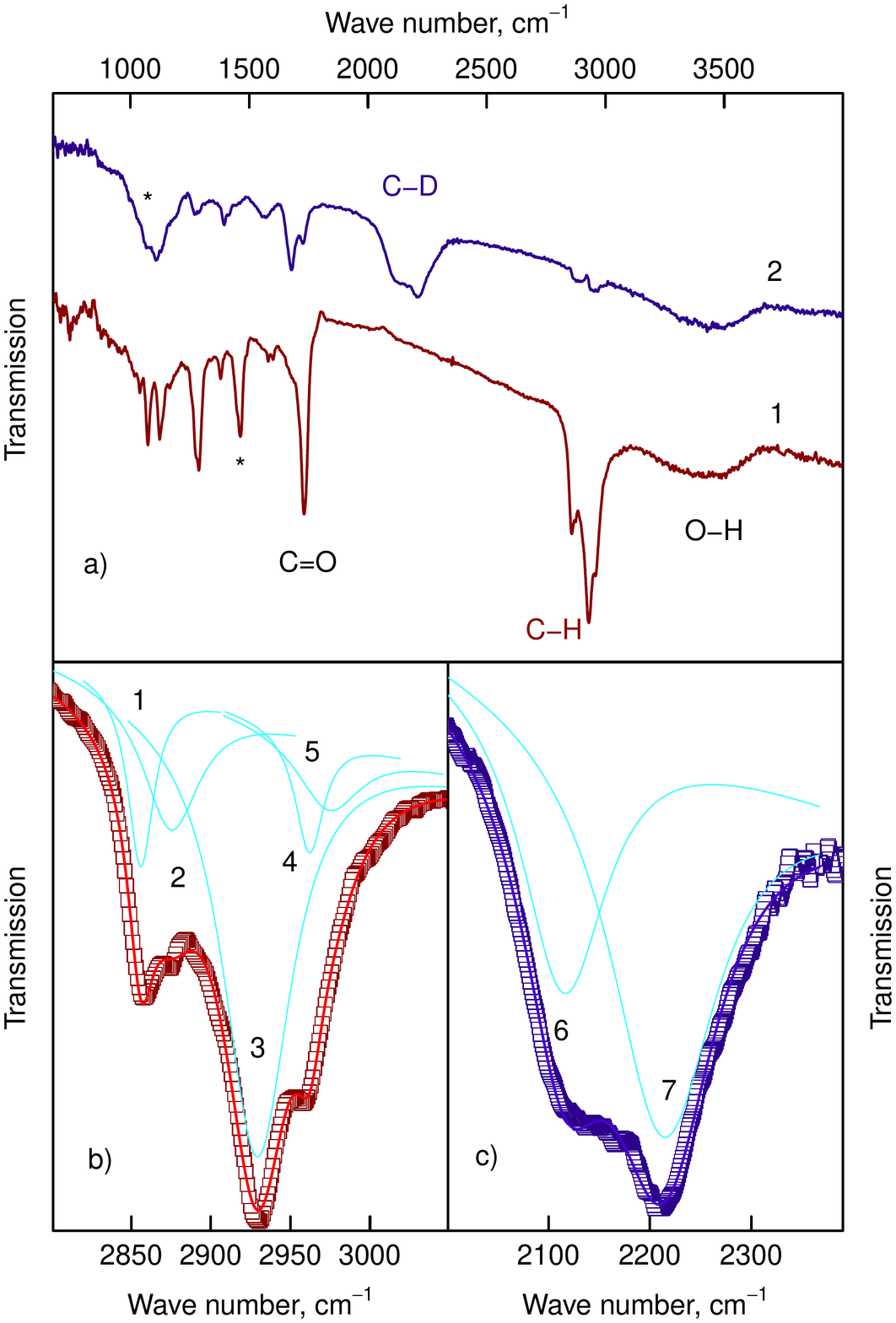}
\caption{ a) IR spectra of samples  34 (1) and 37 (2). 
b) -c). Deconvolution of absorption band of valence bonds oscillations in samples 34 (C--H,  b) and 37 (C--D,  c). Symbols stand for experimental points,  solid lines --  fitting by sum of lorentzians (see Table.~\ref{t1} for peaks parameters and their attribution). A weak absorption band which appears in the spectral range 2700--3000 cm$^{-1}$ for deuterized sample 37 is possibly due to contamination by aromatic hydrocarbons in the process of sample preparation and/or measurement. }
\label{ir}
\end{figure}

Infrared spectra were studied for several samples with different degrees of crystallization, two such spectra for samples 34 and 37 obtained by the polymerization of C$_6$H$_6$ and C$_6$D$_6$ respectively are shown in Fig.~\ref{ir}. 

\begin{table}
\caption{Positions, full width at half maxima parameter (in cm$^{-1}$) and relative intensity of Lorentzians used for fitting of absorption bands of valence bonds oscillations in samples: 34 (C--H,  rows 1--5) and 37 (C--D, rows 6--7). Last two columns are predicted peak positions and attribution of C--H bonds in hydrogenated amorphous carbon films according to \cite{dischler:ssc83}. }
\label{t1}
\begin{tabular}{cccccc}
\hline
& Exp. Pos. & FWHM &  Rel.  Intensity & Theor.  Pos. & Config. \\ \hline
1 &  2856 &  20 &  9 &  2855 & sp$^3$ CH$_2$ (s) \\
2 & 2875 & 44 & 16 &  2870 & sp$^3$ CH$_3$ (s) \\
3 & 2929 & 52 &  58 & 2915 &  sp$^3$ CH \\
4 & 2962 & 22 &  7 & 2925 &  sp$^3$ CH$_2$ (a)\\
5 & 2974 &  50 & 10 & 2960 &  sp$^3$ CH$_3$ (a) \\ \hline
6 &  2115 &  104 &  37 &       &     \\
7 &  2213 & 129 & 63 &         &     \\\hline
\end{tabular}
\end{table}

The main absorption band, which is due to the valence vibrations of C--H bonds, consists of several peaks. The main peaks of this absorption band for sample 34 are in a wavenumber range 2800--3000~cm$^{-1}$ and, consequently, belong to vibrations of aliphatic bonds or sp$^3$ CH$_x$ bonds, where $1\le x\le 3$. According to \cite{dischler:ssc83}, five such vibrations are possible. The description of the absorption band by five Lorentzian curves is shown in Fig.~\ref{ir}~b.

The next strong absorption band that is present in both spectra is a narrow band with a maximum at 1732~cm$^{-1}$, which is attributed to vibrations of C=O bonds. Among other absorption bands in the spectra, the one at 1463~cm$^{-1}$ (marked by asterisk on spectrum 1 in Fig.~\ref{ir}) is certainly attributed to the deformation vibrations of C--H bonds in CH$_2$ and/or CH$_3$ groups, because the substitution of deuterium for hydrogen leads to disappearance of this band and to increase of the band at 1107~cm$^{-1}$ (marked by asterisk on spectrum 2 in Fig.~\ref{ir}) caused by deformation vibrations of CD$_2$ and/or CD$_3$ groups. 

Finally, a wide absorption band in the range of 3200--3600~cm$^{-1}$ is due to the valence vibrations of O--H bonds. The main contribution to its intensity comes from water molecules adsorbed on the surface of particles of the materials under study. The presence of the water absorption band in spectra means that the samples are hydrophilic in contrast to initial benzene, graphite or, e.g., polyethylene.

Thus, the main hydrogen group in sample 34 is the CH group (as in graphane), but the fraction of CH$_2$ and CH$_3$ groups is also noticeable. Since the elemental analysis indicates that the C:H ratio in sample 34 is close to unity, the sample contains some carbon atoms that are not connected to hydrogen atoms. These atoms in the sample under study can be the atoms of carbonyl groups. The appearance of multiple carbon-carbon bonds or diamond-like fragments can also be expected. 

The manifestation of C--D bonds and multiple carbon-carbon bonds is also present in the absorption band of valence vibrations of C--D bonds in sample 37 (Fig.~\ref{ir}~c). This absorption band is fairly wide and poorly structured, so two Lorentzian curves are sufficient to describe it. The center of curve 1 is between the sp$^3$ CD$_2$ and sp$^2$ CD peaks, according to attribution made in \cite{dischler:ssc83}. At the same time, the center of the most intense curve 2 is at the intermediate position between aliphatic and aromatic  sp$^2$ CD peaks. Thus, sample 37 contains a noticeable fraction of aromatic bonds comparable to that present in poorly crystallized C--H samples obtained at lower pressures or temperatures. It is remarkable that the infrared spectra of amorphous polymers obtained earlier under compression at room temperature\cite{pruzan:jcp90,gauthier:hpr92} have a wide unstructured absorption band in the range of 2900--3100~cm$^{-1}$, which indicates a large fraction of aromatic bonds and a low structural ordering of these materials. 

Thus, the main hydrogen groups in samples with a high degree of crystallization are C--H(D) groups connected by aliphatic bonds. 

\begin{figure}
\includegraphics[height=0.8\textheight]{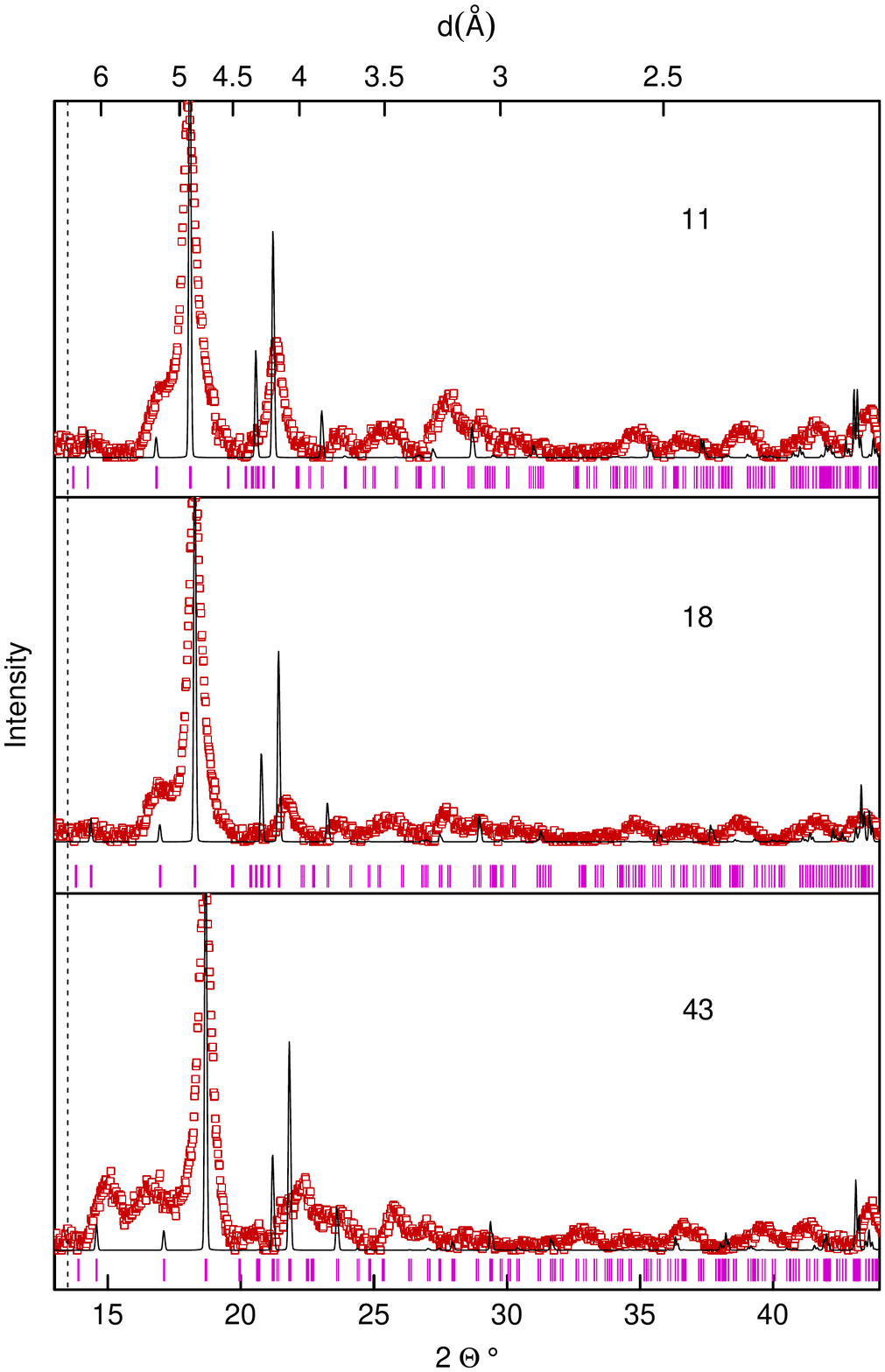}
\caption{Powder X-ray diffraction patterns after subtraction of amorphous baseline of three  graphane samples obtained under different synthesis conditions. Symbols stand for experimental data points, solid lines are simulated XRD patterns and ticks mark positions of reflections allowed by the crystal symmetry. Dashed lines mark the position of the strongest (1 0 0) reflex of nanothread structure according to Ref.~\cite{fitzgibbons:nm14}. }
\label{m1}
\end{figure}

Several possible basic types of graphanes and combinations based on them were considered\cite{wen:pnas11,artyukhov:jpca10,rohrer:prb11,he:pssr12,sahin:wir15,zhou:nrl14}. For samples with a high degree of crystallization, X-ray diffraction data were analyzed. After the removal of an amorphous halo from powder X-ray diffraction patterns, a quite clear crystal diffraction pattern is recovered (Fig.~\ref{m1}). Though it still has quite large full width at half maxima parameters (which is obviously caused by poor crystallization) and cannot be treated by direct crystal structure determination methods, it still enables one to make some guesses about the structure of this material taking into account results (bond lengths and possible interplanar distances) obtained in the previous density functional theory calculations\cite{wen:pnas11,artyukhov:jpca10,rohrer:prb11,he:pssr12,sahin:wir15,zhou:nrl14} .

The main diffraction peak in this picture (d~=~4.75--4.9~\AA) is due to the interplanar distance between hydrocarbon sheets and it almost coincides with the previously predicted distances. Its position depends on the synthesis conditions: shorter interplanar distances appear at higher pressures and temperatures. In view of the purely van der Waals interaction between hydrocarbon sheets, the inclusion of these sheets into the amorphous but totally covalently bonded hydrocarbon network formed under the same high-pressure conditions can exert a strain large enough to produce such variation of this length.

Beside that the main peak in Fig.~\ref{m1} has several left and right satellites, which are obviously caused by the strong corrugation of the hydrocarbon sheet. Four basic types of graphane sheets (A, B, C, and D according to nomenclature from\cite{wen:pnas11}) are flat enough; consequently, the first satellite at higher angles even in the most corrugated sheet D would occur only at angles about 27\degree, which corresponds to the typical in-plane distance between undulations of about 3.2~\AA. Moreover, theoretical considerations predict the absence of any peaks below 18.5\degree (this implies that an interplanar distance of 4.8~\AA\ is the largest interatomic length in this structural model). Both statements obviously contradict experimental observations. 

\begin{figure}
\includegraphics[width=\textwidth]{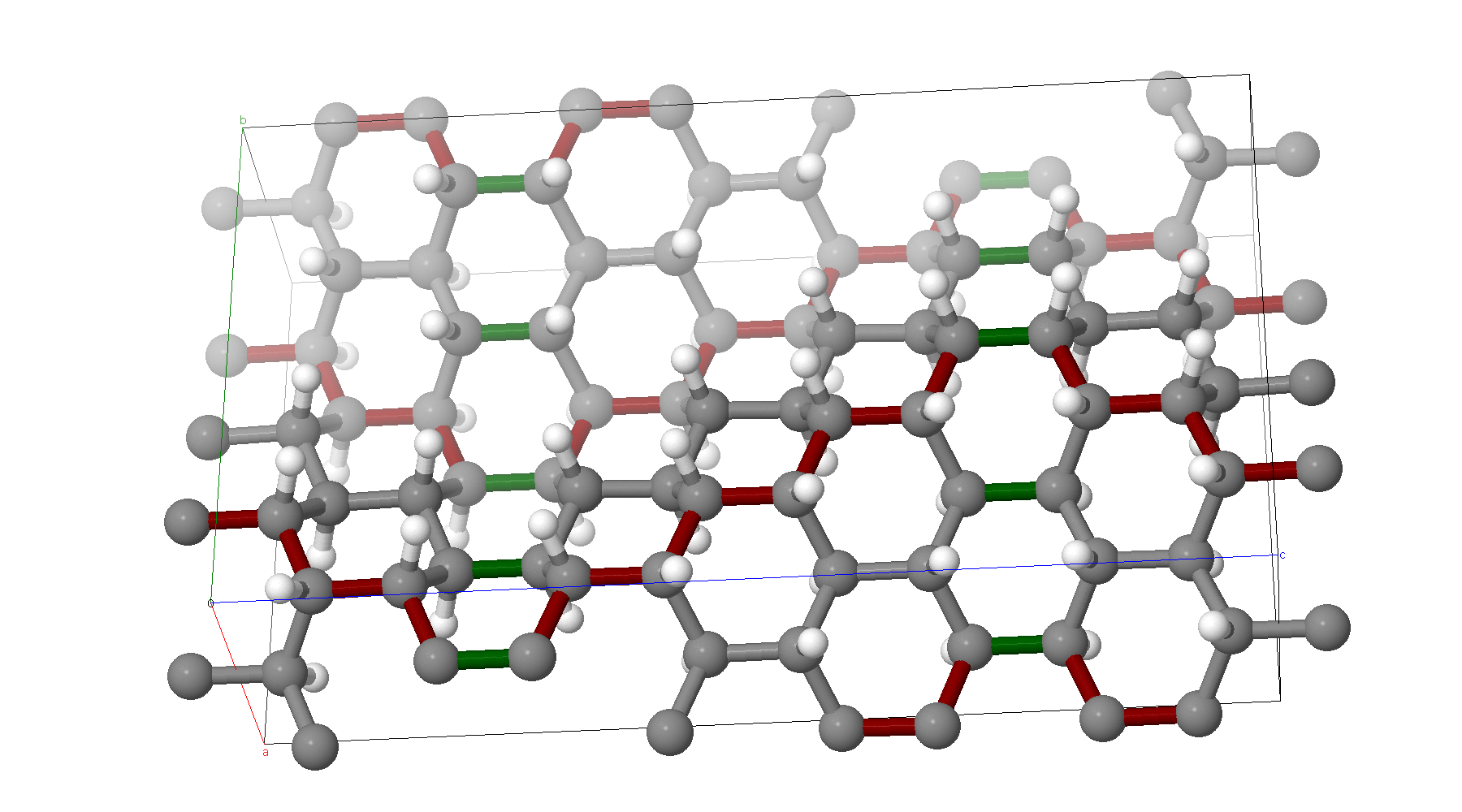}
\caption {Perspective view of single hydrocarbon sheet proposed here as a simplistic structural model. Grey balls stand for carbon atoms, white ones -- for hydrogen. Green bonds represent lonsdaleite type bonds, and red ones mark the directions (``ridges and valleys'' or B-type bonds) where A-type graphane sheet (grey network) is broken. The actual unit cell of the crystal consists of two such sheets shifted by a half bond length along the $c$ axis.}
\label{m2}
\end{figure}

To overcome these challenges, we propose a structural model (Fig.~\ref{m2}), which despite its simplicity catches the most spectacular features of the experimentally observed diffraction pattern. It is very similar to tricycle graphane proposed earlier\cite{he:pssr12}, but it also includes additional defects produced by lonsdaleite-type bonds. It can be regarded as patches of graphane A stitched together in two directions by B-type graphane strips and lonsdaleite (C- or D-type) strips. The respective periodicity of B- and lonsdaleite-type defects suggests the name (3-cycle-4-step) graphane for this graphane sheet, so original tricycle graphane (without any lonsdaleite bonds) would be 3-cycle-1-step. The symmetry of the crystal structure with one such a sheet in the unit cell is $Pbcm$ with $a$~=~4.8~\AA, $b$~=~9.2~\AA, and $c$~=~17.2~\AA. However, the real structure includes another type of defects, namely, stacking faults, caused by the alternative shift of neighboring hydrocarbon sheets about one half of the C--C distance (0.78~\AA) along the $c$ axis, reducing the steric interaction between hydrogen atoms. This implies the doubling of the period along the $a$ axis ($a$~=~9.6~\AA) accompanied by a change in symmetry to the space group $Pbca$.

Hypothetically, one can propose similar structures with a varying number of steps and cycles. Although the introduction of lonsdaleite bonds into the carbon structure generally increases its internal energy, this energy penalty in comparison to diamond bonds is no more than about 10~meV per bond. However the diamond and lonsdaleite-type bonds in this family of graphane sheets are arranged in such a way that they do not produce any additional strain in the crystal lattice.

It should be noted that previous theoretical consideration\cite{wen:pnas11} predicted that pure B- and C-type (lonsdaleite-like) graphane sheets should be thermodynamically stable at pressures about 10 and 20~GPa, respectively. Therefore, under the synthesis conditions used in our experiments (pressures around 10~GPa), it can be conjectured that B-type ``fractures'' of A-type graphene sheets formed as thermodynamically stable at high-pressure conditions can be readily recovered in ambient pressure in a metastable form. On the other hand, we suppose that the orientation of a C--H bond relative to the C--C bonded graphane sheet provides an energy barrier high enough to ``freeze'' kinetically produced lonsdaleite-type bonds (so to say ``accidentally'' obtained under synthesis conditions in the process of growth of graphane sheets) in an even more energetically unfavorable (but still metastable) structure under ambient conditions. It is interesting to note that an increase in the synthesis temperature and pressure (Fig.~\ref{m1}) leads to the appearance of additional peaks in the range of 20\degree--40\degree, which in our opinion is caused by a finer undulation of graphane sheets. Clearly, there are three different energy scales, which produce three types of faults along three orthogonal directions in this hydrocarbon structure.

The density of the proposed lattices (1.75--1.85~g~cm$^{-3}$) is noticeably higher than the experimental values (1.5--1.6~g~cm$^{-3}$). A low macroscopic density is obviously due to a large fraction of the amorphous phase and a high degree of defectiveness of the crystal lattice. 

The inclusion of nitrogen in pyridine results in a significant change in the parameters $b$~=~10.8~\AA\ and $c$~=~16.2~\AA\ at an almost unchanged interplanar distance of 4.8~\AA. Such deviations as compared to pure graphanes are due to the distortion of the angles between C-N bonds.

\begin{figure}
\includegraphics[width=\textwidth]{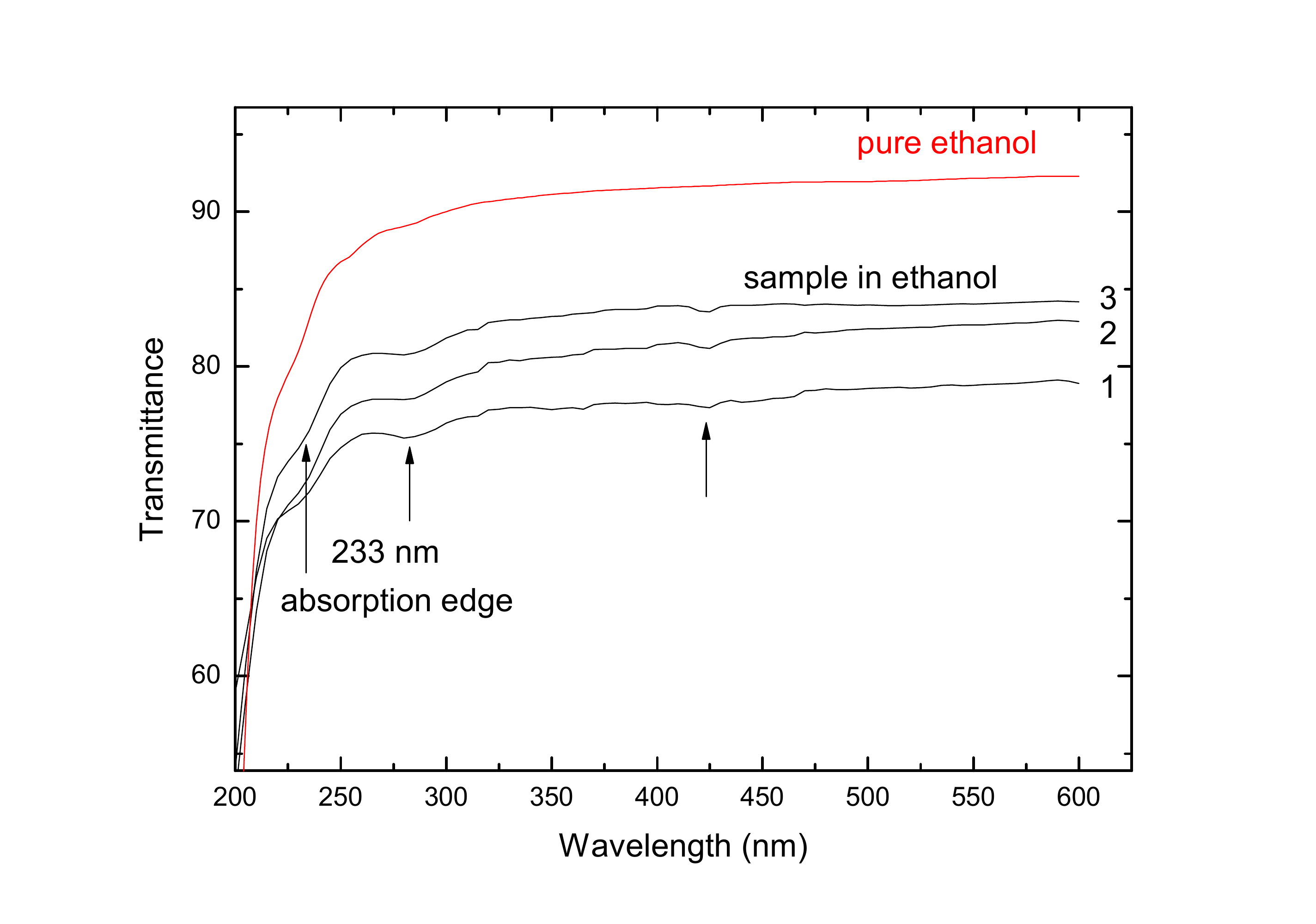}
\caption{Transmission spectra used for optical gap measurements. Baseline of pure ethanol and signal from samples with different concentrations of powdered graphane are shown.}
\label{gap}
\end{figure}

The transmission spectra in optical and UV ranges for samples with a high degree of crystallization are shown in Fig.~\ref{gap}. These spectra includes a small absorption maximum at 2.6--2.9~eV, which is obviously due to intraband absorption, that is responsible for the orange color of graphane samples. One can note that the orange color has been observed previously not only for polymerized benzene samples\cite{pruzan:jcp90,gauthier:hpr92,cansell:jcp93},  but also for other oligomers  fabricated by high pressure-high temperature treatment of different polycyclic hydrocarbons\cite{chanyshev:c15}. Besides there is strong absorption maximum at 3.9--4.1~eV. Finally, there is a smeared absorption edge in the range of 5.1--5.4~eV, which is probably corresponds to the optical gap. The theoretically evaluated optical gap\cite{wen:pnas11,sahin:wir15,zhou:nrl14} in graphanes lies in the range of 3--4~eV  and is likely underestimated. More detailed and sofisticated calculations predicted the values 4.6--5.3~eV for the optical gap\cite{lebegue:prb09,kalicky:jctc13}.

\begin{figure}
\includegraphics[width=\textwidth]{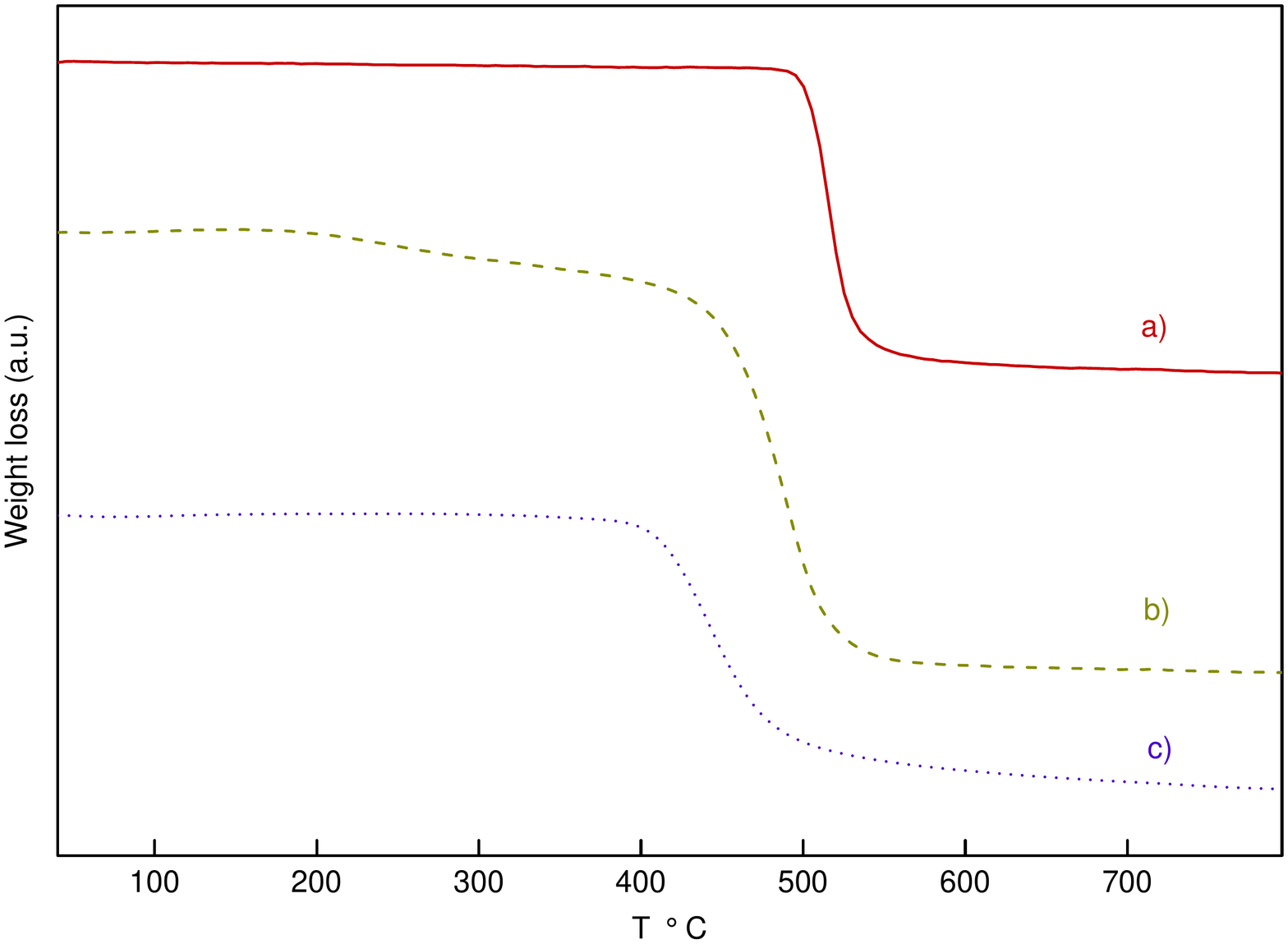}
\caption{Weitgt loss of synthesized samples in the process of decomposition caused by heating with rate  10 K/min. a) -- well-crystallized graphane, b) -- amorphous hydrocarbon phase, c) -- polymeric pyridine phase. The curves are shifted along vertical axis for clarity.}
\label{dta}
\end{figure}

The thermal stability of the samples at normal pressure was studied in the argon atmosphere. Hydrogen leaves graphane samples with the highest degree of crystallization in the narrow interval of 510--540\degree~C, amorphous polymers in the wider range of 420--520 \degree~C, and pyridine-based polymers in the narrow interval of 420--460\degree~C (Fig.~\ref{dta}). A lower thermal stability of pyridine-based polymers correlates well with their stability at high pressures. It is noteworthy that the amorphous phases previously obtained from benzene at room temperature\cite{pruzan:jcp90,gauthier:hpr92} have a significantly lower thermal stability of 200--400\degree~C, which is probably due to incomplete polymerization and a large number of weakly bound elements in the structure.

Since the samples were large enough, their elastic properties could be studied by the ultrasonic method and the microhardness was measured by the Vickers method. The polymerized benzene samples with a high degree of crystallization were primarily studied. For various samples, the bulk modulus $B$ is in the interval of 30--37~GPa, shear modulus $G$ lies in the range of 15--18~GPa, and the Poisson coefficient varies in the interval of 0.2--0.25. The Poisson coefficient is close to values for polycrystalline graphite, whereas the bulk and shear moduli are twice as large as the respective values for graphite. The microhardness of these samples is 1--1.5~GPa, which is also twice as large as the values for polycrystalline graphite. We note that the previous estimate $B$~=~80~GPa obtained in \cite{pruzan:jcp90,gauthier:hpr92} for the bulk modulus of polymerized amorphous benzene is apparently erroneous. 

\begin{figure}
\includegraphics[width=\textwidth]{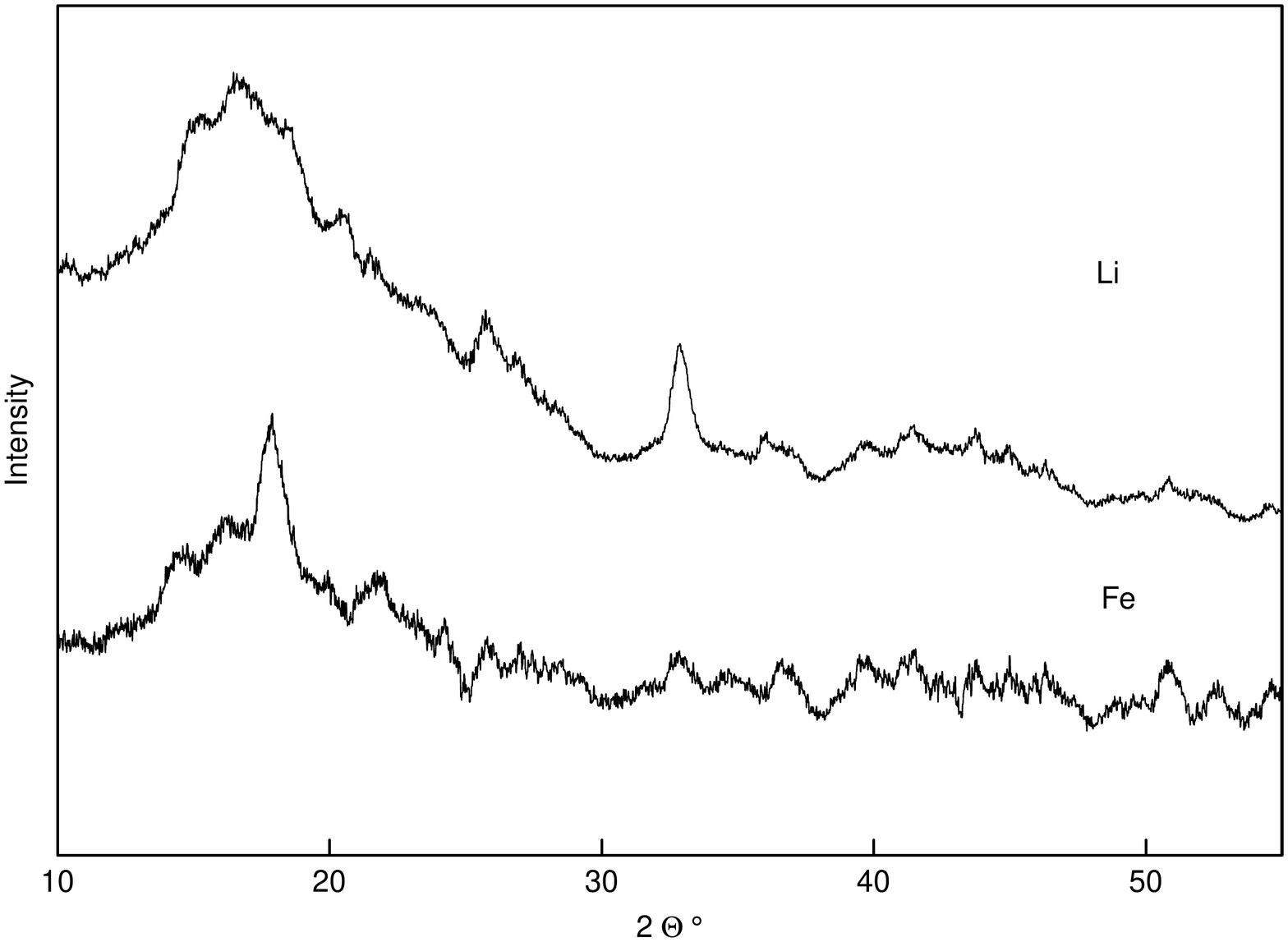}
\caption{XRD patterns of samples doped with Li and Fe.}
\label{dope}
\end{figure}

Benzene and pyridine are good solvents for many metalorganic and boron organic compounds. The thermobaric treatment of such solutions can provide graphane-based structures doped with metals or boron. Fig.~\ref{dope} shows the diffraction spectra for the samples obtained from benzene solutions with 3\% ferrocene (C$_{10}$H$_{10}$Fe) and with 15\% butyllithium (C$_4$H$_9$Li). It is seen that a partially crystallized graphane-based structure can remain at the introduction of a significant fraction of metal atoms (0.5\% for Fe and 2.5\% for Li). The study of the magnetic and electron transport properties of these heavily doped graphanes is now in progress. 

\section{Conclusions}
To summarize, it has been demonstrated that the polymerization of benzene and its analogs simultaneously at superhigh pressures and high temperatures can provide bulk multilayer graphanes. The resulting samples are the best crystallized among all known organic polymers and are the first crystalline polymers with a quasi-two-dimensional layered structure. Bulk samples have the highest density (1.57~g~cm$^{-3}$) for all compounds with C:H=1:1 stoichiometry. 

In future, the degree of crystallization of graphanes can be obviously improved by using higher pressures (10--20 GPa) and by increasing the time of thermobaric treatment from hours to days. The application of higher pressures, along with the use of other aromatic compounds (tetrahydrothiophene, pyrrole) and mixtures of aromatic substances with acetylene, will possibly allow the synthesis of three-dimensionally connected structures with a thermal stability close to graphanes\cite{kondrin:pccp15,kondrin:acb16}. 

Doping of graphanes and other polymerized structures with metals and boron is possible through the solution of metalorganic compounds and borohydrides in the initial molecular liquids. Doped polymers can have extraordinary superconducting, magnetic, and optical properties. 

Although the necessity of high pressures sharply limits the possible size of samples, our results can be suggestion for chemists that these materials, which are more energetcally favourable than benzene, can probably be synthesized by other methods, e.g., catalytic synthesis. We recall that the possibility of polymerization of ethylene was also demonstrated for the first time with the use of high pressures and the method of catalytic synthesis of polyethylene was found only two decades later.

P.S. When this article was under preparation, it was reported in \cite{antonov:c16} that weakly ordered graphane-leke materials with stoichiometry close to 1:1 were obtained by the hydration of a disordered graphite nanopowder at a high pressure. The degree of ordering of the materials obtained in \cite{antonov:c16} is much lower than that in our samples. Nevertheless, the possibility of hydration of carbon materials to the necessary stoichiometry is important for a subsequent search for ways of obtaining large volumes of graphanes.

\ack
Authors are idebted to A.V. Gulyutin, E.L. Gromnitskaya,  S.G. Lyapin,  N.F. Borovikov, M.V. Trenihin, T.B. Shatalova, D.A. Lemenovsky, F.S. Yel'kin  for technical assistance and  thank V.E. Antonov, A.R. Oganov, S.M. Stishov for helpful discussions. Authors are grateful to Russian Science Foundation (Grant No. 14-22-00093) for financial support. 

\section*{References}
\bibliography{local,graph}

\begin{thebibliography}{10}

\bibitem{geim:nm07}
A.~K. Geim and K.~S. Novoselov.
\newblock {The rise of graphene}.
\newblock {\em Nature Materials}, 6:183--191, 2007.

\bibitem{dreyer:acie10}
Daniel~R. Dreyer, Rodney~S. Ruoff, and Christopher~W. Bielawski.
\newblock From conception to realization: An historial account of graphene and
  some perspectives for its future.
\newblock {\em Angewandte Chemie International Edition}, 49(49):9336--9344,
  2010.

\bibitem{sofo:prb07}
Jorge Sofo, Ajay Chaudhari, and Greg Barber.
\newblock {Graphane: A two-dimensional hydrocarbon}.
\newblock {\em Phys. Rev. B}, 75(15):153401, Apr 2007.

\bibitem{sluiter:prb03}
Marcel H.~F. Sluiter and Yoshiyuki Kawazoe.
\newblock Cluster expansion method for adsorption: Application to hydrogen
  chemisorption on graphene.
\newblock {\em Phys. Rev. B}, 68(8):085410, Aug 2003.

\bibitem{elias:s09}
D.~C. Elias, R.~R. Nair, T.~M.~G. Mohiuddin, S.~V. Morozov, P.~Blake, M.~P.
  Halsall, A.~C. Ferrari, D.~W. Boukhvalov, M.~I. Katsnelson, A.~K. Geim, and
  K.~S. Novoselov.
\newblock {Control of Graphene's Properties by Reversible Hydrogenation:
  Evidence for Graphane}.
\newblock {\em Science}, 323(5914):610--613, 2009.

\bibitem{wen:pnas11}
Xiao-Dong Wen, Louis Hand, Vanessa Labet, Tao Yang, Roald Hoffmann, N.~W.
  Ashcroft, Artem~R. Oganov, and Andriy~O. Lyakhov.
\newblock {Graphane sheets and crystals under pressure}.
\newblock {\em Proceedings of the National Academy of Sciences},
  108(17):6833--6837, 2011.

\bibitem{artyukhov:jpca10}
Vasilii~I. Artyukhov and Leonid~A. Chernozatonskii.
\newblock {Structure and Layer Interaction in Carbon Monofluoride and Graphane:
  A Comparative Computational Study}.
\newblock {\em The Journal of Physical Chemistry A}, 114(16):5389--5396, 2010.

\bibitem{rohrer:prb11}
Jochen Rohrer and Per Hyldgaard.
\newblock {Stacking and band structure of van der Waals bonded graphane
  multilayers}.
\newblock {\em Phys. Rev. B}, 83(16):165423, Apr 2011.

\bibitem{wen:jacs11}
Xiao-Dong Wen, Roald Hoffmann, and N.~W. Ashcroft.
\newblock {Benzene under High Pressure: a Story of Molecular Crystals
  Transforming to Saturated Networks, with a Possible Intermediate Metallic
  Phase}.
\newblock {\em Journal of the American Chemical Society}, 133(23):9023--9035,
  2011.

\bibitem{sahin:wir15}
H.~Sahin, O.~Leenaerts, S.~K. Singh, and F.~M. Peeters.
\newblock Graphane.
\newblock {\em Wiley Interdisciplinary Reviews: Computational Molecular
  Science}, 5(3):255--272, 2015.

\bibitem{zhou:nrl14}
Chao Zhou, Sihao Chen, Jianzhong Lou, Jihu Wang, Qiujie Yang, Chuanrong Liu,
  Dapeng Huang, and Tonghe Zhu.
\newblock Graphene's cousin: the present and future of graphane.
\newblock {\em Nanoscale Research Letters}, 9(1):26, 2014.

\bibitem{savini:prl10}
G.~Savini, A.~C. Ferrari, and Feliciano Giustino.
\newblock {First-Principles Prediction of Doped Graphane as a High-Temperature
  Electron-Phonon Superconductor}.
\newblock {\em Phys. Rev. Lett.}, 105(3):037002, Jul 2010.

\bibitem{sofer:n12}
Hwee~Ling Poh, Filip Sanek, Zdenek Sofer, and Martin Pumera.
\newblock High-pressure hydrogenation of graphene: towards graphane.
\newblock {\em Nanoscale}, 4(22):7006--7011, 2012.

\bibitem{sofer:n14}
Zdenek Sofer, Ondrej Jankovsky, Petr Simek, Lydie Soferova, David Sedmidubsky,
  and Martin Pumera.
\newblock {Highly hydrogenated graphene via active hydrogen reduction of
  graphene oxide in the aqueous phase at room temperature}.
\newblock {\em Nanoscale}, 6(4):2153--2160, 2014.

\bibitem{bashkin:jetpl04}
I.~O. Bashkin, Vladimir~Evgen'evich Antonov, A.~V. Bazhenov, I.~K. Bdikin,
  D.~N. Borisenko, E.~P. Krinichnaya, A.~P. Moravsky, A.~I. Harkunov, Yu.~M.
  Shul'ga, Yu.~A. Ossipyan, and E.~G. Ponyatovsky.
\newblock Thermally stable hydrogen compounds obtained under high pressure on
  the basis of carbon nanotubes and nanofibers.
\newblock {\em JETP Letters}, 79(5):226--230, 2004.

\bibitem{kovacic:joc63}
Peter Kovacic and Fred~W. Koch.
\newblock Polymerization of benzene to p-polyphenyl by ferric chloride.
\newblock {\em The Journal of Organic Chemistry}, 28(7):1864--1867, 1963.

\bibitem{ceppatelli:jcp00}
Matteo Ceppatelli, Mario Santoro, Roberto Bini, and Vincenzo Schettino.
\newblock Fourier transform infrared study of the pressure and laser induced
  polymerization of solid acetylene.
\newblock {\em The Journal of Chemical Physics}, 113(14):5991--6000, 2000.

\bibitem{pruzan:jcp90}
Ph. Pruzan, J.~C. Chervin, M.~M. Thi\'ery, J.~P. Iti\'e, J.~M. Besson, J.~P.
  Forgerit, and M.~Revault.
\newblock {Transformation of benzene to a polymer after static pressurization
  to 30 GPa}.
\newblock {\em The Journal of Chemical Physics}, 92(11):6910--6915, 1990.

\bibitem{gauthier:hpr92}
M.~Gauthier, J.~C. Chervin, and Ph. Pruzan.
\newblock Polymerization of benzene and thiophene at high pressure: What is the
  reaction path?
\newblock {\em International Journal of High Pressure Research},
  9(1-6):300--304, 1992.

\bibitem{cansell:jcp93}
Fran\c{c}ois Cansell, Denise Fabre, and Jean-Pierre Petitet.
\newblock {Phase transitions and chemical transformations of benzene up to 550
  C and 30 GPa}.
\newblock {\em The Journal of Chemical Physics}, 99(10):7300--7304, 1993.

\bibitem{ciabini:jcp02}
Lucia Ciabini, Mario Santoro, Roberto Bini, and Vincenzo Schettino.
\newblock High pressure reactivity of solid benzene probed by infrared
  spectroscopy.
\newblock {\em The Journal of Chemical Physics}, 116(7):2928--2935, 2002.

\bibitem{schetino:csr07}
Vincenzo Schettino and Roberto Bini.
\newblock Constraining molecules at the closest approach: chemistry at high
  pressure.
\newblock {\em Chem. Soc. Rev.}, 36:869--880, 2007.

\bibitem{citroni:pnas08}
Margherita Citroni, Roberto Bini, Paolo Foggi, and Vincenzo Schettino.
\newblock Role of excited electronic states in the high-pressure amorphization
  of benzene.
\newblock {\em Proceedings of the National Academy of Sciences},
  105(22):7658--7663, 2008.

\bibitem{ciabini:nm07}
Lucia Ciabini, Mario Santoro, Federico~A. Gorelli, Roberto Bini, Vincenzo
  Schettino, and Simone Raugei.
\newblock {Triggering dynamics of the high-pressure benzene amorphization}.
\newblock {\em Nat. Mater.}, 6(1):39--43, Jan 2007.

\bibitem{bini14}
Roberto Bini and Vincenzo Schettino.
\newblock {\em Materials under extreme conditions: molecular crystals at high
  pressure}.
\newblock World Scientific, 2013.

\bibitem{zhuravlev:prb10}
Kirill~K. Zhuravlev, Katrina Traikov, Zhaohui Dong, Shuntai Xie, Yang Song, and
  Zhenxian Liu.
\newblock Raman and infrared spectroscopy of pyridine under high pressure.
\newblock {\em Phys. Rev. B}, 82:064116, Aug 2010.

\bibitem{fanetti:jcp11}
Samuele Fanetti, Margherita Citroni, and Roberto Bini.
\newblock Structure and reactivity of pyridine crystal under pressure.
\newblock {\em The Journal of chemical physics}, 134(20):204504, 2011.

\bibitem{shinozaki:jcp14}
Ayako Shinozaki, Koichi Mimura, Hiroyuki Kagi, Kazuki Komatu, Naoki Noguchi,
  and Hirotada Gotou.
\newblock Pressure-induced oligomerization of benzene at room temperature as a
  precursory reaction of amorphization.
\newblock {\em The Journal of Chemical Physics}, 141(8), 2014.

\bibitem{ceppatelli:jcp03}
Matteo Ceppatelli, Mario Santoro, Robert Bini, and Vincenzo Schettino.
\newblock High pressure reactivity of solid furan probed by infrared and raman
  spectroscopy.
\newblock {\em The Journal of chemical physics}, 118(3):1499--1506, 2003.

\bibitem{pruzan:jcp92}
Ph. Pruzan, J.~C. Chervin, and J.~P. Forgerit.
\newblock Chemical and phase transformations of thiophene at high pressures.
\newblock {\em The Journal of Chemical Physics}, 96(1):761--767, 1992.

\bibitem{fitzgibbons:nm14}
Thomas~C. Fitzgibbons, Malcolm Guthrie, En~shi Xu, Vincent~H. Crespi,
  Stephen~K. Davidowski, George~D. Cody, Nasim Alem, and John~V. Badding.
\newblock {Benzene-derived carbon nanothreads}.
\newblock {\em Nature Materials}, 14:43--47, 2015.

\bibitem{stalgorova:iet96}
O.~V. Stal'gorova, E.~L. Gromnitskaya, D.~R. Dmitriev, and F.~F. Voronov.
\newblock {Ultrasonic piezometer for the 0-2.0 GPa pressure and 77-300 K
  temperature range}.
\newblock {\em Instruments and Experimental Techniques}, 39(39):880--884, 1996.

\bibitem{dischler:ssc83}
B.~Dischler, A.~Bubenzer, and P.~Koidl.
\newblock Bonding in hydrogenated hard carbon studied by optical spectroscopy.
\newblock {\em Solid State Communications}, 48(2):105 -- 108, 1983.

\bibitem{he:pssr12}
Chaoyu He, C.~X. Zhang, L.~Z. Sun, N.~Jiao, K.~W. Zhang, and Jianxin Zhong.
\newblock Structure, stability and electronic properties of tricycle type
  graphane.
\newblock {\em physica status solidi (RRL): Rapid Research Letters},
  6(11):427--429, 2012.

\bibitem{chanyshev:c15}
Artem~D. Chanyshev, Konstantin~D. Litasov, Anton~F. Shatskiy, Yoshihiro
  Furukawa, Takashi Yoshino, and Eiji Ohtani.
\newblock Oligomerization and carbonization of polycyclic aromatic hydrocarbons
  at high pressure and temperature.
\newblock {\em Carbon}, 84(0):225 -- 235, 2015.

\bibitem{lebegue:prb09}
S.~Leb\`egue, M.~Klintenberg, O.~Eriksson, and M.~I. Katsnelson.
\newblock {Accurate electronic band gap of pure and functionalized graphane
  from GW calculations}.
\newblock {\em Phys. Rev. B}, 79(24):245117, Jun 2009.

\bibitem{kalicky:jctc13}
Franti\^sek Karlick\`y and Michal Otyepka.
\newblock {Band Gaps and Optical Spectra of Chlorographene, Fluorographene and
  Graphane from G0W0, GW0 and GW Calculations on Top of PBE and HSE06
  Orbitals}.
\newblock {\em Journal of Chemical Theory and Computation}, 9(9):4155--4164,
  2013.

\bibitem{kondrin:pccp15}
Mikhail~V. Kondrin and Vadim~V. Brazhkin.
\newblock {Diamond monohydride: the most stable three-dimensional hydrocarbon}.
\newblock {\em Phys. Chem. Chem. Phys.}, 17(27):17739--17744, 2015.

\bibitem{kondrin:acb16}
M.~V. Kondrin, Y.~B. Lebed, and V.~V. Brazhkin.
\newblock {Structure and topology of three-dimensional hydrocarbon polymers}.
\newblock {\em Acta Crystallographica Section B}, 72(4):634--641, 2016.

\bibitem{antonov:c16}
V.E. Antonov, I.O. Bashkin, A.V. Bazhenov, B.M. Bulychev, V.K. Fedotov, T.N.
  Fursova, A.I. Kolesnikov, V.I. Kulakov, R.V. Lukashev, D.V. Matveev, M.K.
  Sakharov, and Y.M. Shulga.
\newblock {Multilayer graphane synthesized under high hydrogen pressure}.
\newblock {\em Carbon}, 100:465 -- 473, Apr 2016.

\end{thebibliography}

\end{document}